\documentclass[a4paper]{aa}
\usepackage{amsmath}
\usepackage{color}
\usepackage{graphicx}
\usepackage{amssymb}
\usepackage[authoryear]{natbib}

\newcommand{\CI}{\hbox{C~{\sc i}}}

\makeatletter

\authorrunning{Nehm\'e et al.}
\titlerunning{Modeling of diffuse molecular gas}

\makeatother
\begin{document}

\title{Modeling of diffuse molecular gas applied to HD~102065 observations}

\author{Cyrine Nehm\'e \inst{1}, Jacques Le Bourlot\inst{1}, 
Fran\c{c}ois Boulanger \inst{2}, 
Guillaume Pineau des For\^ets\inst{2,5} \\
\and C\'ecile Gry\inst{3,4}}

\offprints{Jacques Le Bourlot}

\institute{LUTH, UMR 8102, CNRS, Universit\'e Paris 7 
and Observatoire de Paris, Place J. Janssen, 92195 Meudon, France
\and Institut d'Astrophysique Spatiale, UMR8617, CNRS, 
Universit\'e Paris Sud, Bat. 121, 91405 Orsay Cedex, France 
\and Laboratoire d'Astrophysique de Marseille, UMR 6110, CNRS, Universit\'e de Provence, 
38 rue Fr\'ed\'eric Joliot-Curie, 13388 Marseille cedex 13, France
\and European Space Astronomy Center, RSSD,  P O Box 50727, 28080 Madrid, Spain 
\and  LERMA, UMR 8112, CNRS, Observatoire de Paris, 61 Avenue de l'Observatoire,
75014 Paris, France}

\mail{Jacques.Lebourlot@obspm.fr}

\date{Received / Accepted 23-01-2008}

\abstract
{}
{We model a diffuse molecular cloud 
present along the line of sight to the star HD~102065. We compare our
modeling  with observations to 
test our understanding of physical conditions and chemistry in 
diffuse molecular clouds.}
{We analyze an extensive set of spectroscopic observations which characterize the 
diffuse molecular cloud  observed toward HD~102065.
Absorption observations provide the extinction curve, 
${\rm {H}_{2}}$, \CI ~,
${\rm {CO}}$, ${\rm {CH}}$, and ${\rm {CH}^{+}}$ column densities and excitation.
These data are complemented by 
observations of ${\rm {C}^{+}}$, ${\rm {CO}}$ and dust emission. 
Physical conditions are determined using the Meudon PDR
model of UV illuminated gas.}
{We find that all observational results, except column densities 
of $\rm {CH}$, ${\rm {CH}^{+}}$ 
and  ${\rm {H}_{2}}$  in its excited ($J\geq 2 $) levels, are consistent 
with a cloud model implying a Galactic radiation field
($G\,\sim0.4$ in Draine's unit), a 
density of $80\,{\rm {cm}}^{-3}$ and  a temperature (60-80 K) set by the
equilibrium between heating and cooling processes. To account for excited 
($J\geq 2$) ${\rm {H}_{2}}$ levels column densities, an additional component of  
warm ($\rm \sim 250\, K$) and dense ($\rm n_H \geq 10^4 \, cm^{-3}$) gas  within 0.03~pc of the star 
would be required. This solution reproduces the observations only if 
the ortho-to-para  H$_2$ ratio at formation is $\sim 1$.  In view of the extreme 
physical conditions and the unsupported requirement on the ortho-to-para ratio, we conclude
that H$_2$ excitation is most likely to be accounted for by  the presence of warm molecular
gas within the diffuse cloud heated by the local dissipation of turbulent kinetic energy. 
This warm H$_2$ is required to account for the ${\rm {CH}^{+}}$ column density. 
It could also contribute to the CH abundance and explain the inhomogeneity 
of the CO abundance indicated by the comparison of absorption and emission spectra.}
{}


\maketitle

\section{Introduction\label{Sec_Intro}}

Since the pioneering work of \cite{BD77}, observations of diffuse molecular clouds
continue to motivate and challenge efforts to model the thermal balance and chemistry
of interstellar gas illuminated by UV photons. 
Models allow observers to determine physical conditions 
from their data and observations contribute to models by quantifying  physical processes 
of general relevance to studies of matter in space
such as  H$_2$ formation, photo-electric heating, and cosmic ray ionization. 

Many models of well characterized lines of sight have been presented
\citep[e.g. in the last years:][]{Zsargo,LRH,Shaw06}. 
They are successful in reproducing many observables apart 
from some molecular abundances, 
most conspicuously CH$^+$, which points to out-of-equilibrium chemistry. 
This molecular ion, and several of the molecular species commonly 
observed in diffuse molecular clouds such as CH, OH and HCO$^+$ may be produced by  
MHD shocks \citep{DK86,Pineau86,Flower}, and small scale vortices \citep{Joulain98,Falga06}
where H$_2$ is heated by the localized dissipation of the gas turbulent kinetic energy. 
Turbulent transport between 
the cold and warm neutral medium may also significantly impact the chemistry of diffuse clouds \citep{Lesaffre07}. 

Independently of gas chemistry, 
the presence of H$_2$ at higher temperatures than that set by UV and cosmic-rays 
heating of 
diffuse molecular clouds, 
may be probed through observations of the H$_2$ level populations \citep{CCD}.
A correlation between CH$^+$ and rotationally excited H$_2$ 
was found by \cite{Lambert} using Copernicus observations. 
\cite{Falga05} reported the detection of the S(0) to S(3) H$_2$ lines 
in a line of sight towards the inner Galaxy away from star forming regions. They interpret their 
observation as evidence for traces of warm molecular gas in the diffuse interstellar medium. 
But the interpretation of the wealth of H$_2$ observations provided by the FUSE
satellite  is still a matter of debate.  
\cite{GBNPHF} modeled FUSE H$_2$ observations 
of three stars in Chamaeleon using the Meudon Photon Dominated Regions (PDR) model \citep{JGED}.
They 
show that the model
cannot account for H$_2$ column densities in rotational states with $\rm J>2$.   
A larger sample of H$_2$ FUSE observations \citep{Tumlinson,GillShull05,Wakker06},
including 2 of the 3 Chamaeleon lines of sight of \cite{GBNPHF},  
have been analyzed on the basis of model calculations presented by \cite{BTS}. 
Their model, like other 
PDR models, takes into account the formation of H$_2$ on grains, 
its photo-dissociation by absorption of resonant UV photons, radiative transfer 
and vibrational/rotational excitation resulting from collisions, H$_2$ formation and 
UV pumping. 
Unlike the latest PDR models \citep{Shaw05,LNHLR}, the \cite{BTS} model
does not derive the gas temperature from the thermal balance between heating and cooling processes. 
\cite{BTS} instead consider gas temperature as a model parameter independent of the density, 
incident radiation field and cloud shielding (total extinction). 
They conclude that H$_2$ observations cannot be accounted
for with a single isothermal slab of gas. They propose solutions where the data are fitted
with absorption from two physically independent gas layers with distinct temperatures, 
UV illuminations and column densities.  
A cold ($\le 100\, $K) component contributes most of the total H$_2$ column density while 
a warmer ($\sim 200 \, $K) and thinner component with a higher UV field helps populate the 
high J states. For many of
the model combinations considered by \cite{BTS}, the difference in UV field is too low 
to account for the corresponding difference in temperature. In particular, the combinations proposed for the 
nearby diffuse interstellar medium require an additional heating source in the warm component.
For H$_2$ observations towards Galactic 
stars, some of the absorption may arise from gas in the vicinity of 
the star \citep[e.g.][]{Boisse05}.

In a companion paper \citep{NCI}, we presented a multi-wavelength 
study combining spectroscopic UV, optical, IR and 
radio observations of the interstellar matter, along the line of sight to the nearby
(170 pc) moderately reddened (E(B-V) = 0.17) star HD~102065. 
Absorption observations provide the column densities of 
${\rm {H}_{2}}$ in the J=0 to 5 states, \CI ~in its three fine structure 
states, ${\rm CO}$ in the J=0 to 2 states, ${\rm CH}$ and
${\rm CH}^{+}$ . They are complemented by  observations 
of the ${\rm C}^{+}$, ${\rm CO(2-1)}$ and $(1-0)$ lines and dust continuum 
in emission. Non-detections of the C$_2$ and CN optical absorption lines 
were transformed into upper limits on column densities using
oscillator strengths  listed by \cite{Gredel91,Gredel93}. 

The HD 102065 line of sight is well suited for detailed modeling of 
physical conditions, chemistry and H$_2$ excitation, because of the large amount of 
available data (Table~\ref{tab:XmodXobs}). 
The molecular fraction is not altered by the presence of warm atomic gas along the line
of sight. Comparison of H~I, H$_2$ and dust extinction indicates that the bulk (~90\%) of the 
column density is accounted for by a diffuse molecular cloud identified on IRAS images.
IRAS observations place a strong constraint on 
the presence of matter close to the star. The abundance of CH$^+$ is high. 
This paper extends the work of \cite{GBNPHF}, where
only the FUSE H$_2$ spectrum of HD~102065 was analyzed, to    
a wider set of observations, using an updated version of the Meudon PDR model \citep{LNHLR}. 
\cite{Kopp00} used this sight line to discuss 
the impact of far-UV extinction on the CO abundance.

The structure of the paper is as follow: Sect.~\ref{A-simple-PDR} presents the PDR model
used to characterize the diffuse molecular cloud and Sect.~\ref{sec:observationsmodelling}
describes the main modeling results. First, a reference model is defined by fitting  
observational constraints together. Second, 
the model is compared with each of the observations to assess the dependence of model predictions
on the values of the physical parameters.  
In Sect.~\ref{sec:warmh2}, we present a detailed attempt to model  H$_2$ excitation with a warmer 
component located close to the star.  Sect.~\ref{sec:Conclusions} presents our conclusions.

\section{Description of the PDR model\label{A-simple-PDR}}

We use a comprehensive model of an interstellar cloud, that describes the state of the gas and 
dust exposed to a radiation field as a function of optical depth . The model 
is one-dimensional and stationary.
It presents several improvements over that previously used by \cite{GBNPHF} and
is described in detail in \cite{LNHLR}.
It computes simultaneously, in an iterative way :
\begin{itemize}
\item \textbf{the UV radiative transfer}, taking into account the continuum absorption
by dust and the line absorption by the lowest levels of ${\rm H}_2$. The UV radiation field can
illuminate the cloud from both sides.
\item \textbf{the thermal balance}, taking into account all relevant heating and cooling
processes. 
\item \textbf{the chemistry}, typically coupling about 100 different species through 800
chemical reactions.
\end{itemize}
\subsection{Model parameters and variables\label{param_var}}

Model parameters (see Table~\ref{model_param}) are kept fixed at values
consistent with typical diffuse
clouds and the measured characteristics of the 
HD~102065 line of sight.
The model  ignores the gas velocity structure and assumes 
that the three low velocity components
discussed by \cite{NCI} make a single homogeneous cloud with a visual 
extinction equal to the observed value to HD~102065.
We look for a best fit model by varying only
the UV radiation field strength G (isotropic and incident
on both sides of the cloud) measured in units of Draine's
radiation field, and the gas density $n_{\rm H}$. Density
is constant throughout the cloud. Temperature
is computed by solving for thermal balance.
The standard Draine's field ($G=1$, \citealp{Draine78})
is equivalent to about 1.6 in units of Habing's field
\citep{Habing68}. 

\begin{table}
\caption{Observational constraints and best model results. Upper part
are constraints used in Fig.~\ref{chi2_o}, 
lower part compares unconstrained observations and results.
Number in parentheses are powers of 10.
\label{tab:XmodXobs}}
\begin{center}\begin{tabular}{cccc}
\hline 
\hline 
 & $X^{\rm mod}$ & $X^{\rm obs}$ & $\sigma_{\rm obs}$\\
\hline
$N({\rm CO})/N({\rm{H}_2}$)        & $1.5\,(-7)$    & $1.6\,(-7)$  & $\pm_{0.15\,(-7)}^{0.2\,(-7)}$\\
$N({\rm \CI })/N_H$     & $5.8\,(-7)$    & $6.0\,(-7)$  & $\pm 1.5 \,(-7) $\\
$N($\CI $^{*}_{J=1})/N($\CI )     & $0.17$         & $0.2$        & $\pm0.07$\\
$N($\CI $^{**}_{J=2})/N($\CI )     & $0.03$         & $0.02$       & $\pm0.01$\\
$f_{\rm H_2}=\frac{2N({\rm H}_{2})}{N({\rm H})+2N({\rm H}_{2})}$ & $0.9$     & $0.69$ & $\pm0.12$\\
$N({\rm{H}_2^o})/N({\rm{H}_2^p})$  & $0.73$         & $0.7$        & $\pm0.12$\\
$I({\rm C}^+)\; ({\rm erg/ s\; cm^{2}\; sr})$  & $2.0\,(-6)$    &
$2.8\,(-6)$  & $\pm0.85\,(-6)$\\
\hline
$N({\rm CH})/N({\rm{H}_2})$        & $8.4\,(-9)$    & $1.85\,(-8)$ & $\pm0.3\,(-8)$\\
$N({\rm CN})/N({\rm{H}_2})$        & $1.2\,(-10)$   & $<1.5\,(-9)$ & \\
$N({\rm C}_2)/N({\rm{H_2}})$       & $3.6\,(-8)$    & $<3.5\,(-8)$ & \\
$N({\rm{CO_{J=0}}})/N({\rm{H}_2})$ & $9.0\,(-8)$    & $9.6\,(-8)$  & $\pm_{1.7\,(-8)}^{1.4\,(-8)}$\\
$N({\rm{CO_{J=1}}})/N({\rm{H}_2})$ & $5.1\,(-8)$    & $6.2\,(-8)$  & $\pm_{1.2\,(-8)}^{1.5\,(-8)}$\\
$N({\rm{CO_{J=2}}})/N({\rm{H}_2})$ & $3.7\,(-9)$    & $<7.3\,(-9)$ & \\
\hline
\end{tabular}\end{center}
\end{table}

\begin{table}[!htbp]
\caption{Fixed model parameters. $\delta_{\rm X}$ are the gas 
phase abundance of atom ${\rm X}$ relative to hydrogen.
\label{model_param}}
\begin{center}\begin{tabular}{ccc}
\hline
\hline
Parameter & Value & Comment\\
\hline
$A_{\rm v}$ & $0.67$ mag & Extinction \\
$\zeta$ & $5\,10^{-17}\,{\rm {s}}^{-1}$ & Cosmic Rays ionization\\
$v_{\rm turb}$ & $2.0\,{\rm {km\, s}}^{-1}$ & Turbulent velocity\\
$R_{\rm v}$ & $3.9$ & Total to selective extinction\\
$\omega$ & $4.2\,10^{-1}$ & Dust albedo\\
$g$ & $6.0\,10^{-1}$ & Dust anisotropy factor\\
$G_{\rm r}$ & $0.57\,10^{-2}$ & Dust to gas mass ratio\\
$\rho_{{\rm g}}$ & $2.59\,{\rm {g\, cm}}^{-3}$ & Dust density\\
$\alpha$ & $3.5$ & Dust size distribution index\\
$a_{{\rm min}}$ & $1\,10^{-7}\,{\rm {cm}}$ & Dust minimum radius\\
$a_{{\rm max}}$ & $1.5\,10^{-5}\,{\rm {cm}}$ & Dust maximum radius\\
$\delta_{\rm C}$ & $1.32\,10^{-4}$ & \\
$\delta_{\rm O}$ & $3.19\,10^{-4}$ & \\
$\delta_{\rm N}$ & $7.50\,10^{-5}$ & \\
$\delta_{\rm S}$ & $1.86\,10^{-5}$ & \\
\hline
\end{tabular}\end{center}
\end{table}

\begin{figure}
\begin{center}\includegraphics[
  width=0.7\columnwidth,
 angle=-90]{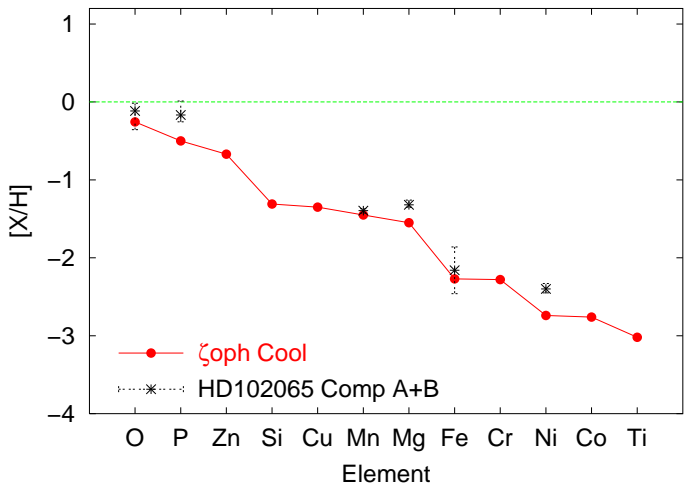}
\end{center}
\caption{Depletion [X/H] in the cool gas towards HD~102065 compared with $\zeta$ Oph values.
Data are taken from Paper I. \label{abun-figure}}
\end{figure}

Element abundances are those measured for $\zeta$ Oph taken from 
\cite{SavSem}. Fig.~\ref{abun-figure} shows that HD 102065 measured
abundances are close to
these reference values (Paper I). The grain distribution is kept fixed.
It is a MRN type one with a power law size distribution $\alpha=-3.5$ with
minimum and maximum sizes $10^{-7}$ and $1.5\times 10^{-5}\,$cm. Although
that distribution does not reproduce the details of the smallest grains
size distribution required to account for the mid-infrared dust emission
\citep{DesBoulPug90, BOUPREG}, it matches the Very Small Grains (VSG
\footnote{By VSG we mean any solid
particle with a typical size smaller than $10^{-6}\,{\rm {cm}}$,
irrelevant of its precise nature.})
to total dust mass ratio of $0.25$.

\subsection{H$_2$ formation equations\label{H2-form-model}}

In the present model, the ${\rm H}_{2}$ 
formation rate on grains is not a free parameter, but is computed locally from 
the combined adsorption of ${\rm H}$ atoms from the gas on grains followed
by recombination and desorption of ${\rm H}_{2}$ molecules:
\begin{eqnarray*}
\rm{H}+dust\rightarrow\rm{H_{ad}}\\
\rm{H_{ad}}+\rm{H_{ad}}\rightarrow\rm{H_2}
\end{eqnarray*}
where $\rm{H_{ad}}$ is an hydrogen atom adsorbed on dust. If only those two processes
 are included and steady state applies, the production of $\rm{H}_2$ is independent of how H atoms 
eventually manage to reach one another and may be computed by:
\begin{equation}
\left.\frac{d[{\rm H}_{2}]}{dt}\right|_{form}={\frac{1}{2}}\;k_{ad}\; [\rm{H}]=
\frac{1}{2}\,s <\sigma n_g>\;\overline{v}_{\rm{H}}\; n(\rm{H})
\label{H2_form_ad}\end{equation}
where $s$ is the sticking coefficient of $\rm{H}$ upon collision,
$\overline{v}_{\rm{H}}$ its mean velocity and
$<\sigma n_g>$ the mean grain cross section per unit volume.

In the full model, other processes may compete with ${\rm H}_2$ formation to remove 
adsorbed hydrogen atoms (photo desorption, reactions on grains, etc...). In this paper, we 
neglect all of those processes, which allows the formation rate to be computed exactly.
 Results from \cite{LEBOU95} show that the mean cross section can be written:
\begin{equation}
<\sigma n_g> = {\frac{3}{4}}\;{\frac{1.4\; \rm{m_H}\; G_r}{\rho_g}} \;
{\frac{1}{\sqrt{a_{\rm min}\times a_{\rm max}}}}
\;n_{\rm{H}}\label{form_H2_sig}
\end{equation}
where G$_r$ is the dust-to-gas mass ratio, $\rho_g$  the density of grain material,
and integration is completed for the entire grain size distribution, using a MRN
distribution \citep{MathisRN}
with an index of $3.5$ from $a_{\rm min}$ to $a_{\rm max}$.
Combining  Eq ({\ref{H2_form_ad}}) and ({\ref{form_H2_sig}})
we can write:
\begin{equation}
 \left.\frac{d[{\rm H}_{2}]}{dt}\right|_{form}=R_f\;\sqrt{T_{\rm gas}}\; s \; n_{H}\;
n({\rm H})\label{H2_form_rate_final}
\end{equation}
where $R_{\rm f}$, the $\rm{H_2}$ formation rate in $\rm{cm^{-3}\;s^{-1}}$, reads:
\begin{equation}
R_{\rm f} = 1.27\times 10^{-20} \; \frac{G_{\rm r}}{\rho_{\rm g}} \;
\frac{1}{\sqrt{a_{\rm min}\times a_{\rm max}}}
\end{equation}

The resulting value of $R_{\rm f}$ is $2.3\times 10^{-17} \; cm^{-3} \, s^{-1}$.
We use a prescription from David Flower (see discussion in \citealp{LNHLR}) for the
sticking coefficient $s$:
\begin{equation}
s = \sqrt{10/T_{\rm gas}({\rm K})}
\end{equation}

The ortho-to-para ratio on formation of H$_2$ is set to a value of three. 
One third (1.5~eV) of the formation energy is transferred into internal
excitation, and is distributed over all H$_2$ energy levels with a Boltzmann distribution
(\citealp{LNHLR}). A second third of the formation energy is converted 
into kinetic energy of H$_2$, and the last third, into grain heating.

\section{Modeling the molecular cloud toward HD102065\label{sec:observationsmodelling}}

\subsection{General modeling results\label{sub:model-results}}

\begin{figure}
\begin{center}\includegraphics[%
  width=0.7\columnwidth,
  angle=-90]{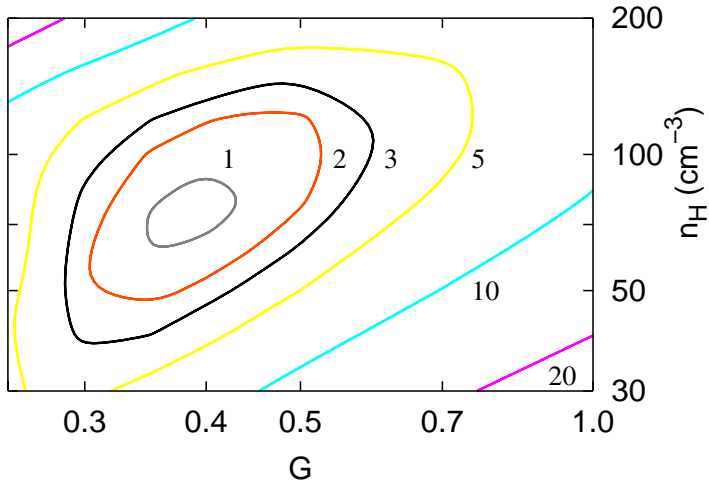}\end{center}
\caption{$\chi^{2}$ contours (Eq~\ref{eq_chi2}) using the top 7 quantities
of Table~\ref{tab:XmodXobs}. The best fit is obtained for $G=0.4$
and $n_{\rm H} = 80\, {\rm cm}^{-3}$. Contours are labeled with
the $\chi^2$ value.\label{chi2_o} }
\end{figure}

To compare observations and model results, we have selected {\emph seven}
quantities for which an observational error bar could be computed:
$N({\rm CO})/N({\rm H}_{2})$, $N({\rm \CI })/N_H$,
$N({\rm \CI ^*_{J=1}})/N({\rm \CI })$,
$N({\rm \CI ^{**}_{J=2}})/N({\rm \CI })$,
$f_{{\rm H}_2}=2N({\rm H}_2)/(N({\rm H})+2N({\rm H}_{2}))$,
$N({\rm H}_2^{\rm o})/N({\rm H}_2^{\rm p})$, and $I({\rm{C}^+})$,
where $N({\rm X})$ is the column density of species ${\rm X}$ over the whole
cloud and ${\rm H}_2^{\rm o}$ and ${\rm H}_2^{\rm p}$ are
ortho- and para-${\rm{H}_2}$ respectively. From these quantities,
we define a $\chi^{2}$ error function by:
\begin{equation}
\chi^{2} = \frac{1}{7}\, \sum_{i=1}^{7} \left(\frac{X_i^{\rm obs}-X_i^{\rm mod}}{\sigma_{\rm obs}}\right)^{2}
\label{eq_chi2}
\end{equation}
where $X_i^{\rm obs}$ is the quantity derived from observations,
$X_i^{\rm mod}$ the same quantity from a model, and $\sigma_{\rm obs}$
the observational uncertainty (see Table~\ref{tab:XmodXobs})
\footnote{Note that $X_i$ need not be dimensionless because normalisation
by $\sigma$ ensure that compatible quantities are summed}.
We computed a grid of models for gas densities ranging from 30 to 200 $cm^{-3}$ and
UV fields $G$ from 0.25 to 1.0. Figure~\ref{chi2_o} shows the
resulting $\chi^{2}$ iso-values. Contours
are  smooth, and limit a well defined minimum where
$\chi^{2}<1$, i.e. where model results are, \emph{in the mean}, closer
to observations than observational uncertainties.

\begin{figure}
\begin{center}\includegraphics[
  width=0.7\columnwidth,
  angle=-90]{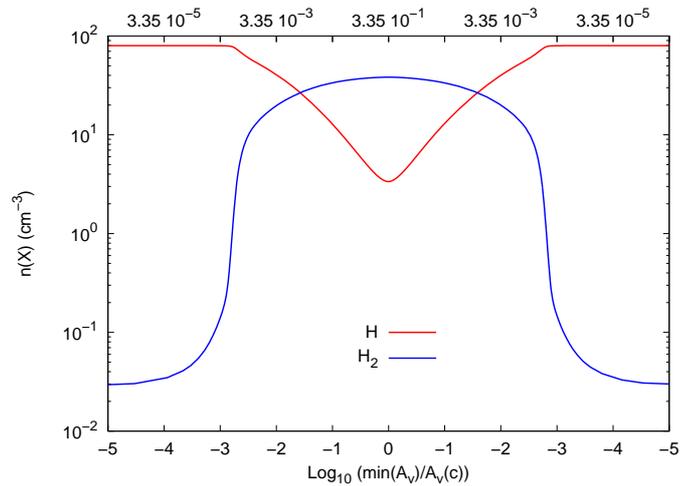}
\end{center}
\caption{${\rm H}$ and ${\rm H}_2$ density profiles for the reference model plotted
versus the extinction from the nearest edge normalized to the central
extinction.}
\label{CoupeHH2}
\end{figure}

\begin{figure}
\begin{center}\includegraphics[
  width=0.7\columnwidth,
  angle=-90]{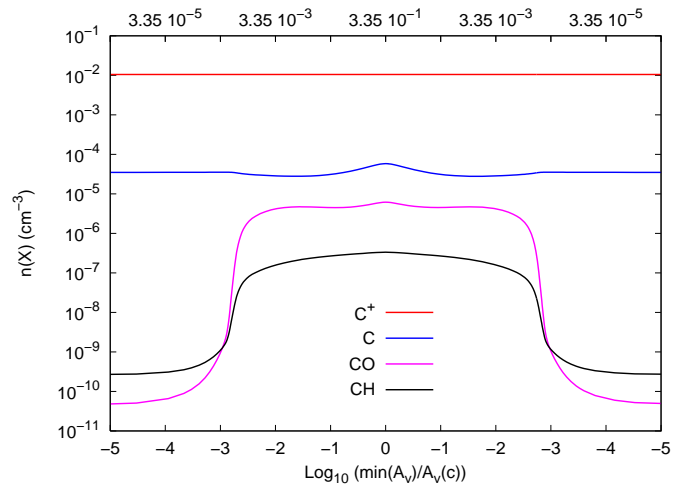}
\end{center}
\caption{${\rm C}^+$, ${\rm C}$, ${\rm CO}$ and ${\rm CH}$ density
profiles for the reference model.
\label{CoupeCOCCH}}
\end{figure}

\begin{figure}
\begin{center}\includegraphics[
  width=0.7\columnwidth,
  angle=-90]{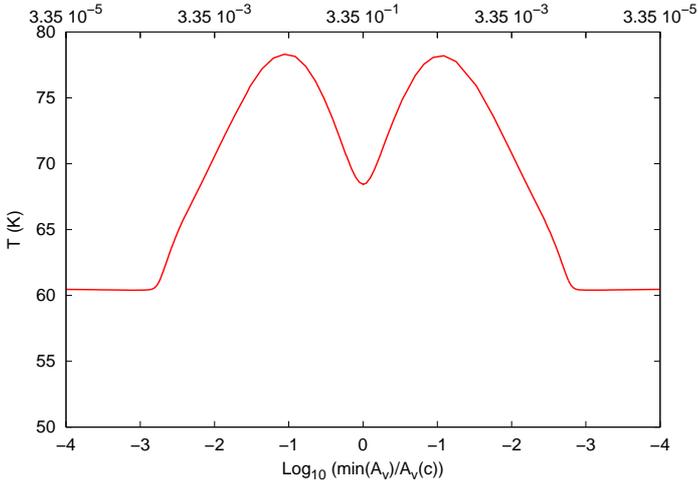}
\end{center}
\caption{Temperature profile for the reference model.
\label{Temp}}
\end{figure}

One can see that the best compromise is reached for a rather low radiation
field ($G=0.4$, which is close to the standard Habing field) and
a mean total proton density of $n_{\rm H}=80\,{\rm{{cm}}}^{-3}$.
In the following we will refer to those values as our reference model.
The reference model results are compared to the observables in Table~\ref{tab:XmodXobs}.
Abundance and temperature profiles are
illustrated on Fig.~\ref{CoupeHH2}, \ref{CoupeCOCCH} and
\ref{Temp}\footnote{The peculiar horizontal scale allows for
logarithmic scaling towards both sides of the cloud}.
The ${\rm H~{\sc I}}$ to ${\rm H}_2$ transition occurs close to the cloud edge at an extinction
of $A_{v}\sim10^{-3}$.
Most of the carbon is in the form of  ${\rm C}^+$ at all depths. The fraction of atomic carbon
\CI ~is constant throughout the cloud. The density profiles of ${\rm CO}$ and ${\rm CH}$
molecules follow that of ${\rm H}_2$. Temperature varies from 60 to 80 K, and
is lowest at the cloud edge. The temperature bumps visible on Fig.~\ref{Temp} are due to
heating by ${\rm H}_2$ formation.

The gas heating rate is $1.8\times 10^{-24}\,{\rm{{erg\, cm}}}^{-3}\,{\rm s}^{-1}$
at the cloud center and comes mainly from photo-electric effect on grains.
In the outer layers where the temperature peaks, photo-electric
heating reaches $3.2\times 10^{-24}\,{\rm{{erg\, cm}}}^{-3}\,{\rm s}^{-1}$
and there is a lower but significant heating due to ${\rm H}_{2}$
formation ($2.2\times 10^{-24}\,{\rm{{erg\, cm}}}^{-3}\,{\rm s}^{-1}$).
Cooling is dominated by ${\rm C}^{+}$, with ${\rm H}_{2}$
contributing up to 20\% around $10^{-2}\, A_{v}$. This leads
to an integrated ${\rm{{C}^+}}$ emissivity 
of $2\times 10^{-6}\,{\rm{{erg\, cm}}}^{-2}\,{\rm s}^{-1}\,{\rm{{sr}}}^{-1}$
in good agreement with the ISO observation.

In the next sections, we discuss each of the observables and their
dependence on the two model parameters n$_{H}$ and $G$.

\begin{figure}
\begin{center}\includegraphics[
  width=0.7\columnwidth,
  angle=-90]{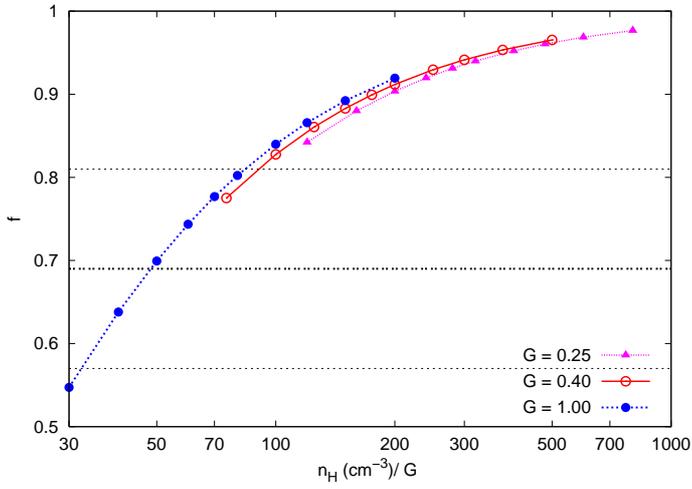}
\end{center}
\caption{${\rm H}_2$ fraction $f_{\rm H_2}$ as a function
of $n_{\rm H}/G$ for $G = 0.25, 0.4\; \&\; 1.0$. The horizontal
lines are the observed value with the error bars.
\label{Coupef}}
\end{figure}

\begin{figure}
\begin{center}\includegraphics[
  width=0.7\columnwidth,
  angle=-90]{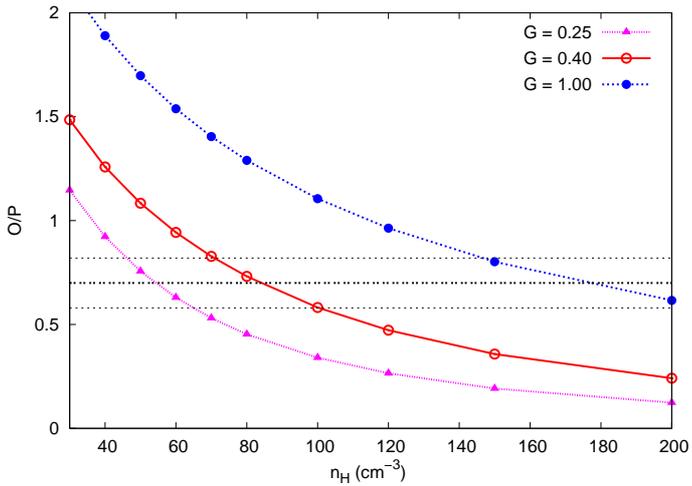}
\end{center}
\caption{Ortho-to-para ratio of ${\rm H}_2$ as a function of $n_{\rm H}$ for three values of $G$.
\label{coupe_osp}}
\end{figure}

\begin{figure}
\begin{center}\includegraphics[
  width=0.7\columnwidth,
  angle=-90]{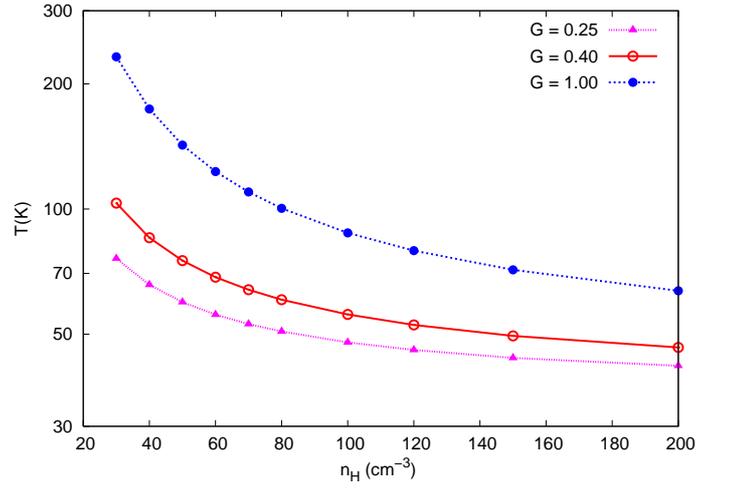}
\end{center}
\caption{Mean gas temperature as a function of $n_{\rm H}$ for three values of $G$.
\label{coupe_temp}}
\end{figure}

\subsection{ ${\rm H}_{2}$ formation rate and molecular fraction\label{sub:form-h2}}

${\rm H}_2$ fraction (f$_{\rm{{H_2}}}$) is plotted versus
density for three values of $G$ on Fig.~\ref{Coupef}. 
For the reference model ($G = 0.4$, $n_{\rm H} = 80\,{\rm{{cm^{-3}}}}$) computed values
for f$_{\rm{{H_2}}}$ are around 0.9,  significantly higher than the observed f$_{H_2}$.

\cite{GBNPHF} have shown that $f_{\rm{H_2}}$, for a given value of $G$, 
depends on the product $n_{\rm H} \times R$, where ${\rm H}_2$ formation
proceeds at a rate $n_{\rm H}\; n({\rm H})\; R$.
We note that their $R$ incorporates both the temperature dependence and the sticking coefficient.
Using our prescription for the sticking coefficient $s$ (section~(\ref{A-simple-PDR})), 
one finds that R is equal to $\sqrt{10}\;R_{\rm f}$ and is   
independent of $T_{\rm gas}$.

More generally, Fig.~\ref{Coupef} shows that the molecular fraction 
for a given total gas column density 
depends on $n_{\rm H} \times R /G$. \cite{GBNPHF}
found that the observed f$_{\rm{{H_2}}}$ is matched for
$n_{\rm H} \times R = 2.3~10^{-15}\; {\rm s}^{-1}$, for a cloud illuminated 
on one side by the Habing field (equivalent to $ G = 0.6$ in Draine units).
Our reference model gives $n_{\rm H}\; \sqrt{10}\; R_{\rm f} = 5.8\; 10^{-15}\, {\rm s}^{-1}$, 
a factor 2.5 higher. This difference explains why, in our model, the molecular fraction is higher than
the observed value.
In the model, the formation of molecular hydrogen is the only destruction path
of adsorbed atomic hydrogen in the present calculation. Other processes
(such as photo desorption) may limit the amount of ${\rm H}$ adsorbed on grains
and lower further the ${\rm H}_2$ formation efficiency.
We consider the agreement between model and observed H$_2$ fraction satisfactory because 
the time scale of H$_2$ formation, a few $ 10^7$yr, is long compared to dynamical time scales
in diffuse clouds. The observed value is thus not expected to accurately 
match the model steady state value. Analysis of a large set of FUSE lines of sight will be required to 
estimate how observed molecular fractions scatter about model steady state values  and 
constrain the ${\rm H}_2$ formation rate more precisely.

\subsection{${\rm H}_2$ excitation\label{sub:h2excit}}

\begin{figure}
\begin{center}\includegraphics[
  width=0.7\columnwidth,
  angle=-90]{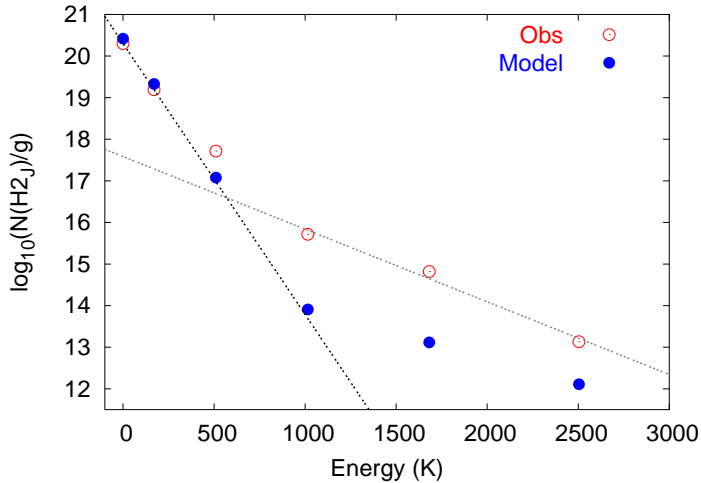}
\end{center}
\caption{Observed H$_2$ excitation diagram and reference model.
Two excitation temperatures are plotted for 
$J=0$ and $1$: $T_{\rm ex}=66\,\rm K$, and for $J=3$ to $5$: $T_{\rm ex}$=248\,$\rm K$.
\label{Excdiag}}
\end{figure}

The ortho-to-para ${\rm H}_2$ ratio, and the mean gas temperature 
are shown in Fig.~\ref{coupe_osp} and \ref{coupe_temp}, and the ${\rm H}_2$ excitation
diagram in Fig.~\ref{Excdiag}.
${\rm{{H_2}}}$ column densities in the $J=0$ and $J=1$  levels are reasonably well reproduced.
The observed ortho to para ratio depends on the gas temperature. If it is at
its equilibrium value (this is a reasonable hypothesis for diffuse clouds where the
time scale for ortho-to-para conversion is shorter than that of H$_2$ formation), 
the $N(J=1)/N(J=0)$ ratio can be considered a direct measure 
of the gas temperature. 
The excitation diagram in Fig.~\ref{Excdiag} implies
an excitation temperature of $66\,{\rm{{K}}}$ (for J$\leq 2$), which indeed is in good 
agreement with the mean gas temperature for the reference model (see Fig.~\ref{Temp}).
 
On the other hand, ${\rm H}_{2}$
column densities  derived from the model for $J>2$ are far below the
observed values. For levels $J\geq3$, we derive an excitation temperature
$T_{{\rm{ex}}}=248\,{\rm{{K}}}$. This high $J$ gas suggests
the existence of a warmer gas component that will be discussed in
Sect.~\ref{warmH2}. We used the model to check that an increase
in the cosmic-rays ionization rate, a possibility considered by  
\citet{Shaw06}, is not a solution. Multiplying this rate by a factor
20 to a value of $10^{-15}\,$s$^{-1}$ 
increases the model column densities in the J=3 and 4 levels by 
less than a factor of 2. 

\subsection{Carbon abundance and excitation\label{sub:carbun-exci}}

\begin{figure}
\begin{center}
\includegraphics[width=0.7\columnwidth,angle=-90]{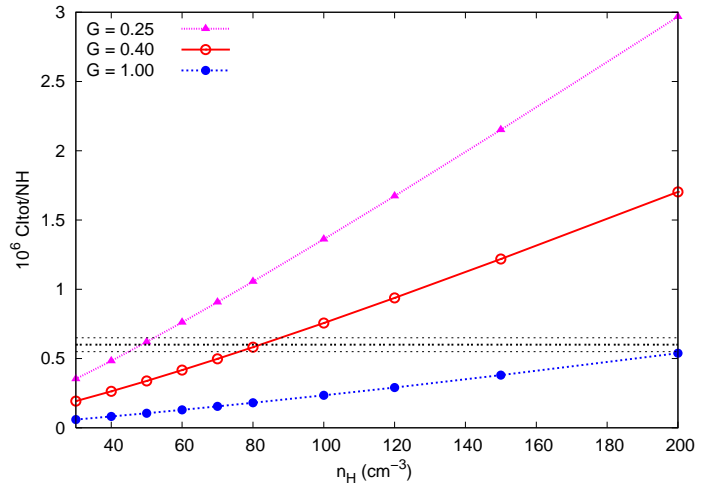}
\end{center}
\caption{Neutral atomic carbon abundance as a function of the gas density 
$n_{\rm H}$ for three values of the UV radiation field intensity factor $ G $.}
\label{CI_b_CINH}
\end{figure}

\begin{figure}
\begin{center}\includegraphics[
  width=0.7\columnwidth,
  angle=-90]{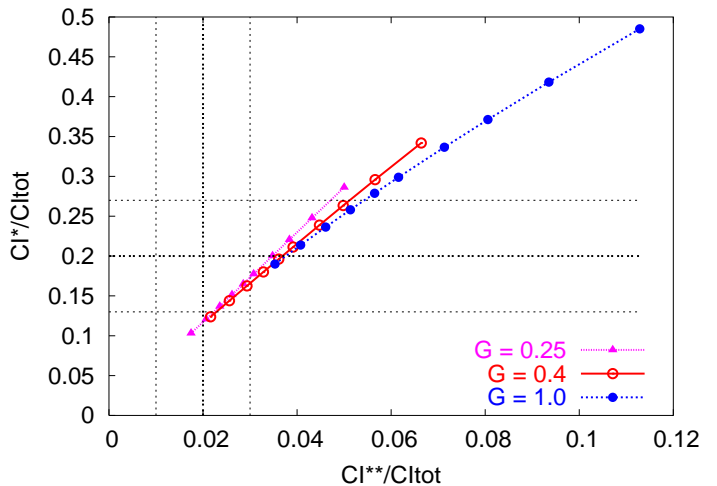}
\end{center}
\caption{Fraction of excited \CI ~(J=1 and 2 fine structure levels: 
\CI $^{*}$ and 
\CI $^{**}$) for various $n_{\rm H}$ and $G$.
Dashed lines are the observations with their
3-$\sigma$ uncertainties. Along each line, the densities 
increase from left to right from 30 to $200\,$cm$^{-3}$ as in Fig.~\ref{CI_b_CINH}. }
\label{CI_a_CINH}
\end{figure}

The neutral atomic carbon abundance and fine structure excitation, are presented
in Fig.~\ref{CI_b_CINH} and \ref{CI_a_CINH}.
Most of the carbon is in ${\rm C}^+$
and the fraction of carbon in the neutral form \CI ~depends
on both the gas density and the value of $G$. It increases with 
densities and decreases with increasing $G$. Our reference model
is in  good agreement with the observed value.

The neutral atomic carbon excitation in diffuse clouds is often considered as a measure of gas
pressure \citep{JT01}. In our model, this translates into a dependence
on the product $n_{\rm H} \times G$ with only small differences from one $G$
to another. Our reference model, reproduces well the fraction of $J=1$
fine structure level, but not so well that of $J=2$, which implies n$_{H}\le 50 \, {\rm cm^{-3}}$ for $G=0.4$.
It is often the case that \CI  ~excitation  in diffuse clouds cannot be accounted for by gas
at a single pressure \citep{JT01}. 
However our observations show a lower $ \CI ^{**}/\CI ^*$ ratios than implied by the
model, contrary to \cite{JT01} findings.

\subsection{Molecules\label{molecule}}
\label{Sec:molecules}

\begin{figure}
\begin{center}\includegraphics[%
  width=0.7\columnwidth,
  angle=-90]{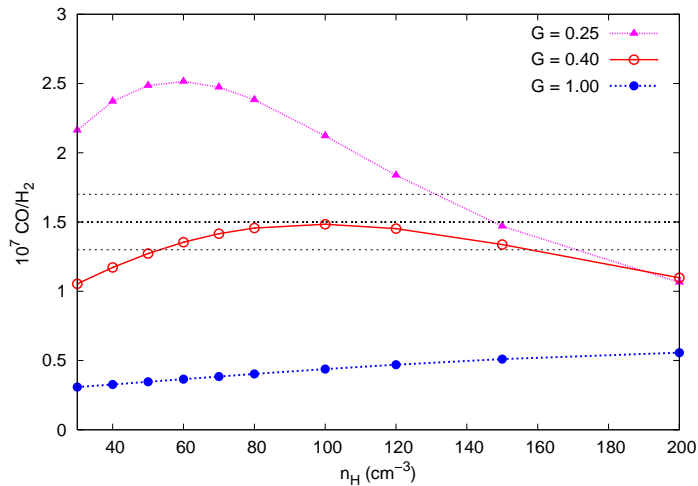}
\end{center}
\caption{${\rm N(CO)}/{\rm N(H_2)}$ variation with density.
All other model parameters are as in Table~\ref{tab:XmodXobs}. The horizontal
lines show the observed values with the error bars \label{rad_dens_coup_CO}.}
\end{figure}

\begin{figure}
\begin{center}\includegraphics[%
width=0.80 \columnwidth] {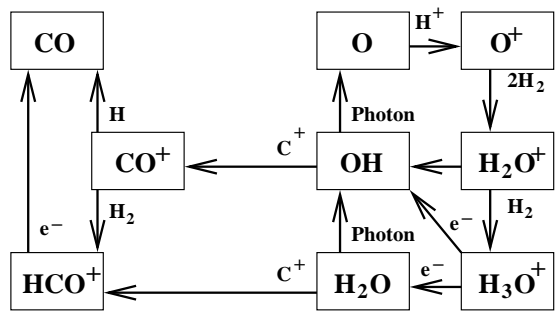}\end{center}
\caption{A schematic illustration of ${\rm CO}$ formation routes in
a low UV radiation field. Destruction is still dominated by Photodissociation\label{COevolution}.}
\end{figure}

Our reference model matches the observed abundance of $\rm CO$
(Fig.~\ref{rad_dens_coup_CO}). 
Table~\ref{tab:XmodXobs} shows that $\rm CO$ excitation is
also well predicted by the model, although it was not used as a constraint.
The rise in the $N({\rm CO})/N({\rm H}_{2})$ ratio at low density
and low radiation field is a chemical effect.
The formation of ${\rm CO}$ is illustrated in Fig.~\ref{COevolution}. 
The efficiency of $\rm CO$ formation depends on the ionization fraction.
Photoionisation is proportional to the gas density $n_{\rm H}$,
while recombination proceeds at a rate that is proportional to the square of that quantity. 
The degree of ionization therefore increases when the density decreases in a
given radiation field. Enhancing the ionization enhances ${\rm{{O}}}^{+}$
formation via charge exchange with ${\rm H}^{+}$ and favors the formation
of $\rm OH$ and $\rm H_2O$. Both molecules
interact with ${\rm C}^+$ and lead either directly or
indirectly to the formation of $\rm CO$. The reaction involving OH is dominant.
Photodestruction of $\rm CO$
is limited by the weakness of the radiation field.

We are cautious about the model interpretation of the CO abundance. It is noticeable that the 
CO column densities in the J=1 and 2 levels, derived from UV spectra, are significantly smaller
by factors 3 and 7, respectively, than the values derived from the emission 
radio spectra \citep{NCI}. These differences indicate that the 
abundance of $\rm CO$ is inhomogeneous, 
which could be accounted for within the PDR model, by introducing clumps with higher density
than the mean value. However, the large CH$^+$ abundance 
favors an alternative
explanation, where a small-scale increase in the
CO abundance, traces the localized contribution of out-of-equilibrium chemistry to its formation.
\cite{Liszt00} have gathered results from UV and radio absorption measurements of CO along 
diffuse interstellar medium lines of sight. They relate the CO abundance to that of 
HCO$^+$ measurements, concluding that CO formation through dissociative recombination of 
HCO$^+$ suffice to account for the CO abundance in diffuse molecular clouds. \cite{Falga06} show
that the observed abundance of HCO$^+$ ($\sim 2 \times 10^{-9}$) cannot be accounted by 
standard PDR chemistry and must be related, like CH$^+$, to warm out-of-equilibrium chemistry.
A significant fraction of CO observed in diffuse molecular clouds 
may thus be a product of out-of-equilibrium chemistry. 

This interpretation links
CO abundance inhomogeneities to the  CH$^+$ chemistry but it is not 
specifically the CH$^+$ rich gas that has an enhanced CO abundance.   
It is not ruled out by the observed velocity difference
between CH$^+$ and CO \citep{NCI}. CO, unlike CH$^+$, 
is observed to be concentrated in the intermediate velocity component C. This component 
may correspond to shielded sections of the cloud where the CO
photo-dissociation rate is reduced. 
In \cite{NCI}, we propose that the line of sight to HD~102065 samples 
material ablated  from the Dcld~300.2-16.9 cloud by a cloud-supernova shock interaction.
In this scenario, the matter flowing out of the cloud is expected to be 
very turbulent \citep{Nakamura06}.

\begin{figure}
\begin{center}\includegraphics[%
  width=0.7\columnwidth,
  angle=-90]{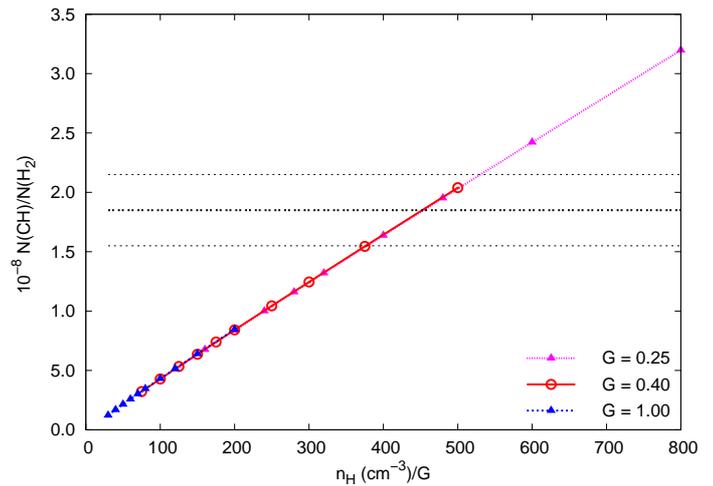}
\end{center}
\caption{${\rm{{N(CH)}}}/{\rm{{N(H_2)}}}$ variation with density and the radiation field. 
 The horizontal lines show the observed values with the error bars. \label{Fig:ch_ch}}
\end{figure}

In Fig.~\ref{Fig:ch_ch}, we show that the CH abundance   
(${\rm{{N(CH)}}}/{\rm{{N(H_2)}}}$) computed by the model depends linearly on the ratio $n_{\rm H}/G$.
The value for our reference model is a factor of two lower than observations.
This mismatch between model and observations, 
might be an additional manifestation of the out-of-equilibrium chemistry, 
as already proposed by e.g. \cite{Zsargo} and \cite{Ritchey06}.
The ``excess'' of $\rm CH$ abundance may be the product of 
$\rm CH^+$ recombination with H$_2$. Finally, we note that the model
is consistent with the upper limits on CN and C$_2$ abundances (Table~\ref{tab:XmodXobs}).

\section{Warm H$_{2}$\label{sec:warmh2}}

\label{warmH2}

The $\rm H_2$ excitation diagram (Fig.~\ref{Excdiag}) shows that the
observed column densities at $J>$2 lie far above 
the single temperature fit to the low $J$ column densities.
This is commonly observed, towards many stars by Copernicus and FUSE.
The model values, which take into account UV pumping and $\rm H_2$ excitation 
upon formation are also roughly an order of magnitude lower than the observations.
\cite{GBNPHF} and others earlier (e.g. \citealp{DK86, Joulain98, Pineau86}) have proposed that this 
excited ${\rm H}_{2}$ traces warm gas in regions
where kinetic energy is dissipated through shocks or vortices. Other authors have 
 proposed that this warm  ${\rm H}_{2}$
is the signature of molecular gas close to the star and
thus exposed to a high UV field (e.g. \citealp{BTS}).  
The presence of CH$^+$ in quantities much larger than predicted by the PDR model favors 
the former explanation \citep{Falga05}. To test this preference,  
we have used the PDR model to quantify the latter possibility 
in the specific case of  HD~102065.

The ratio between the $100 \mu{\rm m}$ brightness in the IRAS images
and the visible extinction towards HD~102065 indicates that most of the
matter is fairly distant from the star \citep{BOUPREG} and does not interact with it. 
The 60 and 100~$\mu$m  images  only show a small
brightness enhancement at the position of HD~102065, point-like
at the IRAS resolution (angular size  $\leqslant$ 5'), and 
corresponding to a small 
fraction (~10$^{-3}$) of the stellar luminosity \citep{BOUPREG}. If the absorbing matter occupies
a solid angle $\theta$ about the star, its UV/visible opacity is $10^{-3}\times 4\pi /\theta$.
Combining the constraint on the source diameter ($< 5'$), and the star distance (170 pc), we derive
an upper limit for the distance, from the star to the absorbing matter of $0.12\; {\rm pc}$. 
HD102065 is a B9IV star with a luminosity $\sim 100\,L_\odot$ and an effective temperature 
of $11300$~K. 
The stellar radiation field intensity is ${G_*= 0.2\times (d/1{\rm pc})^{-2}}$ in Draine units.


For an extinction $A_{\rm v} \sim 10^{-3}$ the gas is molecular only for large densities.
To search for a more realistic solution, we assume that $\theta/4\pi \sim 0.1$.
To quantify with this assumption, the column density of warm H$_2$ that could be associated
with the IRAS far-IR emission, 
we compute a grid of models with constant $A_{\rm v} = 10^{-2}$,
varying the density $n_{\rm H}$ from $10^3$ to $3\times 10^6 {\rm cm^{-3}}$, 
and the distance to the star from $0.1 {\rm pc}$ to $1\; 10^{-2}\; {\rm pc}$
(equivalent to $G_*$ from 20 to $2\times 10^{3}$). 


\begin{figure}
\begin{center}\includegraphics[
  width=0.7\columnwidth,
  angle=-90]{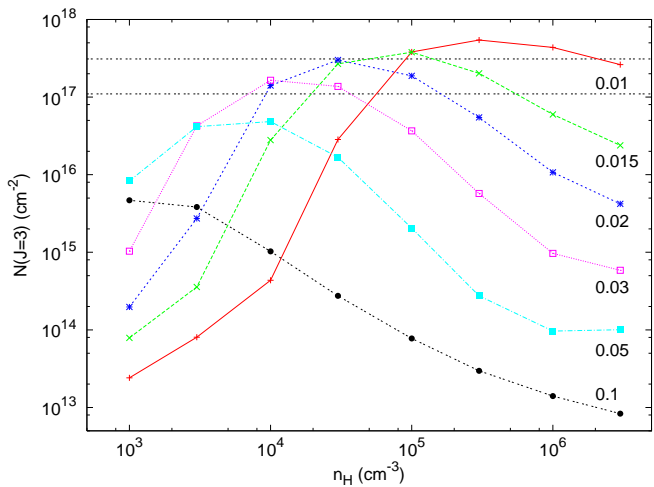}
\end{center}
\caption{${\rm H}_2\; (J=3)$ column densities for an $A_{\rm v}=10^{-2}$ slab
of gas close to the star (labels next to the curves is $d$ in pc).
The two dashed horizontal lines bracket the measured column density with its error bar.
\label{NH2exc_nh}}
\end{figure}

Figure~\ref{NH2exc_nh} shows 
the column densities of ${\rm H}_2\; (J=3)$, as a function of $n_{\rm H}$ for various
distances to the star. As distance decreases, the radiation field increases,  
the H$_2$ photo-dissociation rate rises, and higher 
densities are required to retain hydrogen in its molecular form. The stronger  radiative
pumping populates higher rotational levels, which accounts for the bell
shape of the curves. The same trend occurs at higher 
rotational levels, for which the curve maxima are shifted towards higher densities.

\begin{figure}
\begin{center}\includegraphics[
  width=0.7\columnwidth,
  angle=-90]{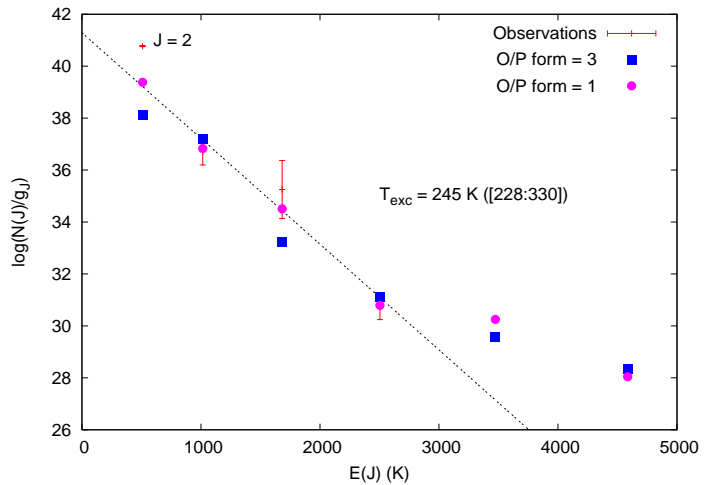}
\end{center}
\caption{${\rm H}_2$ excitation diagram for $n_{\rm H}=~3\,10^4 \, {\rm cm}^{-3}$, 
 $d=0.02\; {\rm pc}$ ($G \sim 300$). The J=4 column density is reproduced only
if the ${\rm H}_2$ ortho to para ratio at formation would be 1 but no theoretical nor
experimental study  support a value much different from 3. 
\label{H2_exc}}
\end{figure}

The observed excitation temperature of $\sim 250\; {\rm K}$ is reproduced for
the correct column densities, in a range of models, as shown on Fig.~\ref{NH2exc_nh}. 
One typical example is illustrated in 
Fig.~\ref{H2_exc}. It can be clearly seen however that the $J=4$ point lags under the $J=3-5$
curve. This is directly linked to the imposed ortho-to-para ratio on the formation
on grains. This is because the high radiation field results in a photodissociation time scale 
which is an order of magnitude larger than the conversion time scale from ortho-to-para 
${\rm H}_2$ through reactive collisions, with either ${\rm H}$ or ${\rm H}^+$.
The only way to reproduce the $J=4$ column density
is to assume a formation ortho-to-para ratio that is different from the statistical
equilibrium value of 3. Fig.~\ref{H2_exc} shows that an initial ratio of 1
provides robust results. We are however unaware of any theoretical or experimental result 
that would support such an initial ratio.

We conclude that a pure steady-state PDR model is able to reproduce 
the excitation of ${\rm H}_2$ at $J>2$,  observed towards HD~102065,
with two {\it ad hoc} assumptions: the presence of warm and dense gas 
close to the
star at a pressure $\rm > 10^6\; K \, cm^{-3}$, and an H$_2$ ortho-to-para 
ratio, at formation, of 1.
We checked  the possibility that the warm H$_2$ may be
in a circumstellar disk. Upper limits on ${\rm H}_2$ column densities from the $\beta$
Pictoris debris disk, are several orders of magnitude lower than the
HD~102065 values \citep{Lecav01}. The warm ${\rm H}_2$ column densities
observed towards HD~102065 are comparable to those detected from circumstellar disks 
about the young AeBe Herbig stars HD~100546 and HD~163296 \citep{Lecav03},
but these disks are traced by strong near to mid-infrared emission that is not observed in HD~102065.

\begin{figure}
\begin{center}\includegraphics[
  width=\columnwidth,
 angle=0]{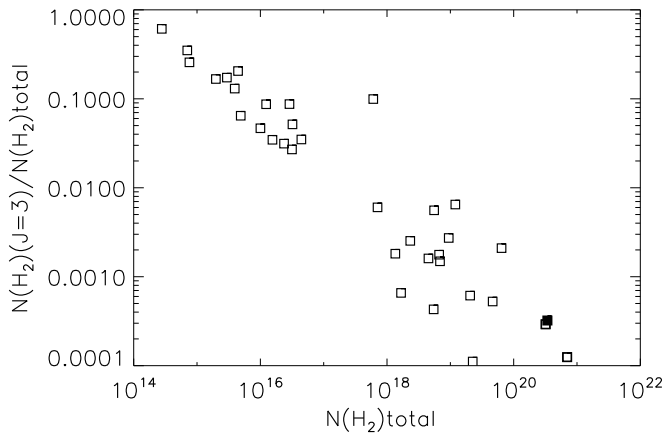}
\end{center}
\caption{Fraction of ${\rm H}_2$ in the J=3 level versus 
the total H$_2$ column density. The data are taken from 
observations of Galactic H$_2$ towards extragalactic sources as analyzed by \cite{GillShull05}.
The three data points at $\rm N(H_2) \, > \, 10^{20} \, cm^{-3}$ correspond to
the three Chamaeleon stars of \cite{GBNPHF} including HD~102065 (filled square). 
\label{n3ebvh2}}
\end{figure}

\cite{GillShull05} and \cite{Wakker06} analyzed a large set of 
FUSE Galactic H$_2$ detections obtained towards 
extragalactic sources. \cite{Wakker06} showed that H$_2$ excitation, for column density ratios 
between the 4 first J levels, follows a systematic trend with increasing total H$_2$ column density. 
HD~102065, and the two remaining Chamaeleon stars studied by
\cite{GBNPHF}, fall on the same trend extending to yet higher $\rm N(H_2)$ values (Fig.~\ref{n3ebvh2}).  
This agreement indicates that 
H$_2$ excitation observed towards the Chamaeleon stars, fit with observations 
towards extragalactic sources and 
is thus unlikely to be due to matter heated by the stars.
Qualitatively, the trend observed in Fig.~\ref{n3ebvh2} is in agreement with the fact that the gas temperature 
and the H$_2$ radiative pumping increase, as the
H$_2$ column density decreases. But, 
based on the HD~102065 results, 
we anticipate that a warm out-of-equilibrium H$_2$  component is required to quantitatively reproduce the data.
We conclude that the presence of some  
H$_2$ at temperatures higher than the equilibrium
temperature set by UV and cosmic-ray heating 
is a general characteristic of diffuse molecular gas in the Solar Neighborhood.
Modeling of the wide sample of FUSE Galactic H$_2$ measurements, is 
required to statistically quantify the fraction of out-of-equilibrium 
H$_2$ gas. This warm molecular gas traces the local dissipation 
of turbulent kinetic energy. It is 
the dissipation of turbulence that 
creates the non-equilibrium chemistry discussed in 
Sect.~\ref{Sec:molecules}.

\section{Conclusions\label{sec:Conclusions}}

We have gathered a wide set of observations characterizing the diffuse 
molecular cloud, observed towards the star HD~102065. These observations 
provide independent constraints on physical conditions, which
are analyzed with the latest version of the Meudon PDR code. 

A single density ($n_{\rm H} = 80\; {\rm cm}^{-3}$) slab of gas, bathed 
in a low radiation field, accounts for most observations
(molecular fraction, gas temperature inferred from the $\rm N(H_2,J=1)/N(H_2,J=0)$, 
cooling in the C$^+$ line,  \CI  ~abundance and excitation). 
This model provides a physical and chemical reference upon which
more elaborate interpretations can be developed to account for 
observables not reproduced by the model, namely the column densities 
of $\rm {CH}$, ${\rm {CH}^{+}}$ and  ${\rm {H}_{2}}$  in its excited ($J\geq 2 $) levels.

We consider the possibility that high J H$_2$ is associated
with matter close to the star. We place constraints on
this possibility using IRAS data on dust emission.  We find a solution 
where dense ($\rm n_H \geq 10^4 \, cm^{-3}$) gas  with a high pressure ($\rm p/k \, > \, 10^6\; K \, cm^{-3}$)
would be located within 0.03 pc of the star. This solution requires the presence of high pressure
gas close to the star and that the 
H$_2$ ortho-to-para ratio at formation is 1. Such a departure from the statistical
expectation is supported by no theoretical or experimental study. 

The H$_2$ excitation observed towards HD~102065 fits
with the general trend observed from FUSE Galactic H$_2$ observations towards extragalactic
sources. We conclude that H$_2$ excitation in the $\rm J>2$ levels 
observed towards HD~102065, is unlikely to be due to matter heated by the star, but 
it is more likely characteristic of H$_2$ in the Solar Neighborhood diffuse ISM.
Our work supports earlier studies that proposed that H$_2$ excitation  in the $\rm J>2$ levels, 
traces the presence of warm H$_2$, 
heated by the localized dissipation, in space and time, of turbulent kinetic energy within diffuse molecular 
clouds. The warm H$_2$ is physically associated with the bulk of the  molecular gas, 
traced by the J=0 and 1 H$_2$ absorptions, and 
is required to account for the
${\rm {CH}^{+}}$ column density. 
In addition, the warm H2 could contribute to the CH abundance, and the inhomogeneity 
of the CO abundance, as indicated by the comparison of absorption and emission spectra.

This paper outlines a framework for modeling the large number of Galactic H$_2$ measurements derived from 
FUSE extragalactic observations, which cover 6 orders of magnitudes in  H$_2$ column densities.
In diffuse clouds, the time scale of H$_2$ formation, a few $ 10^7$yr, 
is long compared to dynamical time scales. Observed values of the gas fraction in H$_2$ 
are thus not expected to match steady-state values. Modeling of the available data will
quantify how observed molecular fractions scatter about model values,  and 
better constrain the ${\rm H}_2$ formation rate than the present study. 
Modeling of these observations  will, in addition, statistically
quantify the presence of warm H$_2$, heated by localized dissipation of kinetic energy.

\end{document}